\documentclass[tradiabstract]{aa}  

\usepackage{graphicx}
\usepackage{txfonts}
\usepackage{longtable}
\usepackage{lscape}

\newcommand{\Mjup}{$\mathrm{M_{J}}$}

\newcommand{\Msun}{$\mathrm{M_\odot}$}
\newcommand{\Rjup}{$\mathrm{R_{J}}$}

\newcommand{\Rsun}{$\mathrm{R_\odot}$}
\newcommand{\Lsun}{$\mathrm{L_\odot}$}

\newcommand{\etal}{et al.}
\newcommand{\Kepler}{\emph{Kepler}}
\newcommand{\Corot}{\emph{CoRoT}}
\newcommand{\kms}{km~s$^{-1}$}
\newcommand{\ms}{m~s$^{-1}$}

\begin{document}

	\title{SOPHIE velocimetry of \Kepler\ transit candidates}
	\subtitle{I. Detection of the low-mass white dwarf \textbf{KOI 74b}\thanks{Based on observations collected with the SOPHIE spectrograph on the 1.93-m telescope at Observatoire de Haute-Provence (CNRS), France, by the SOPHIE consortium (program 10A.PNP.CONS).}}

  	\author{D.~Ehrenreich\inst{1}, A.-M.~Lagrange\inst{1}, F.~Bouchy\inst{2,3}, C.~Perrier\inst{1}, G.~H\'ebrard\inst{2,3}, I.~Boisse\inst{2}, X.~Bonfils\inst{1}, L.~Arnold\inst{3}, X.~Delfosse\inst{1}, M.~Desort\inst{1}, R.~F.~D\'\i az\inst{2,3}, A.~Eggenberger\inst{1}, T.~Forveille\inst{1}, C.~Lovis\inst{4}, C.~Moutou\inst{5}, F.~Pepe\inst{4}, F.~Pont\inst{6}, N.~C.~Santos\inst{7,8}, A.~Santerne\inst{3,5}, D.~S\'egransan\inst{4}, S.~Udry\inst{4} \& A.~Vidal-Madjar\inst{2}}

   	\offprints{D.~Ehrenreich}

   	\institute{
     Laboratoire d'Astrophysique de Grenoble, Universit\'e Joseph Fourier, CNRS (UMR 5571), BP 53, Grenoble CEDEX 9 \email{david.ehrenreich@obs.ujf-grenoble.fr}     
     \and     
     Institut d'Astrophysique de Paris, Universit\'e Pierre et Marie Curie, CNRS (UMR 7095), 98 bis, boulevard Arago 75014 Paris, France     
     \and     
     Observatoire de Haute-Provence, CNRS/OAMP, 04870 Saint-Michel-l'Observatoire, France
     \and
     Observatoire de Gen\`eve, Universit\'e de Gen\`eve, 51, chemin des Maillettes, 1290 Sauverny, Switzerland
     \and
     Laboratoire d'Astrophysique de Marseille, Universit\'e de Provence, CNRS (UMR 6110), 38, rue Fr\'ed\'eric Joliot-Curie, 13388 Marseille CEDEX 13, France
     \and
     School of Physics, University of Exeter, Exeter, EX4 4QL, UK
     \and
     Centro de Astrof\'\i sica, Universidade do Porto, rua das Estrelas, 4150-762 Porto, Portugal
     \and
     Departamento de F\'\i sica e Astronomia, Faculdade de Ci\^encias, Universidade do Porto, Portugal
   	}

   	\date{}
 
	\abstract{The \Kepler\ mission has detected transits and occultations of a hot compact object around an early-type star, the Kepler Object of Interest KOI 74. The mass of this transiting object was photometrically assessed in a previous study using the presence of the relativistic beaming effect (so-called `Doppler boosting') in the light curve. Our aim was to provide a spectroscopic validation of this pioneering approach. We measured the radial velocity variations of the A1V star \object{KOI 74} with the SOPHIE spectrograph at the 1.93-m telescope of the Observatoire de Haute-Provence (France). Radial velocity measurements of this star are challenging because of the high level of stellar pulsations and the few available spectral lines. Using a technique dedicated to early-type main-sequence stars, we measured radial velocity variations compatible with a companion of mass $0.252\pm0.025$~\Msun, in good agreement with the value derived from the \Kepler\ light curve. This work strengthens the scenario suggesting that KOI 74 is a blue straggler orbited by a stellar core despoiled of its envelope, the low-mass white dwarf \object{KOI 74b}.}


	\authorrunning{Ehrenreich \etal}
	\titlerunning{Velocimetric detection of the white dwarf KOI 74b}
	\maketitle
%

\section{Introduction}
\label{sec:intro}
Space missions like \Corot\ (Baglin \etal\ 2006) and \Kepler\ (Borucki \etal\ 2010a) are designed to find transiting Earth-size planets. Transit observations provide the size of the transiting planets (more exactly, the planet-to-star radius ratios). Spectroscopic measurements of the host star's radial-velocity variations are then required to determine the masses of the transiting objects. In addition, velocimetry can rule out grazing transits of binary stars and stellar blends, which could mimic a planetary transit signal. Consequently, there is a definite symbiosis between both techniques, since on the one hand, velocimetry is needed to establish the true nature of the transit candidates, while on the other, transiting systems have an orbital inclination on the sky $i\sim90\degr$, allowing the degeneracy to be lifted on the minimum mass $m_2 \sin i$ provided by velocimetry.

Two puzzling and challenging transiting systems were recently identified among the candidates revealed by the \Kepler\ mission. Rowe \etal\ (2010) present light curves featuring the transits and occultations of Kepler Objects of Interest KOI 74 and KOI 81. Here, we focus on KOI 74, for which the occultation is deeper than the transit.\footnote{The occultations are distinguished from the transits by the absence of limb-darkening effect.} It means that this transiting object has a higher surface brightness than its host star, an A1V star with $T_\mathrm{eff} = 9\,400$~K. The story became even more intriguing when Rowe \etal\ (2010) calculated the radius $r_2$ of this hot object from the transit observations 	as 0.4~Jovian radius (\Rjup).

The depth of the occultation is $(r_2/r_1)^2(F_2/F_1)$, where $F_2$ and $F_1$ are the brightnesses of the companion and the primary star, respectively, and $(r_2/r_1)^2$ is the depth of the transit with $r_2$ and $r_1$ the radii of the companion and its primary star, respectively. It is therefore possible to calculate the surface temperature of the companion, found to be around $13\,000$~K (van Kerkwijk \etal\ 2010). This peculiar transiting object is hot and compact; however, a mass estimation is required to unravel its nature further. 

Without any available radial velocities, Rowe \etal\ (2010) tentatively derive a mass for the companion assuming that the $\sim45$-ppm amplitude in the half-period photometric variations exhibited by this system was due to changes in the observable stellar surface area caused by tidal distortions of the host star. The amplitude of these `ellipsoidal' variations is indeed proportional to the mass ratio between the two components of the system (Pfahl, Arras \& Paxton 2008). They find a mass between 0.02 and 0.11~\Msun\ (or 20.9 and 115~\Mjup) for KOI 74b, i.e., in the brown-dwarf or low-mass star domain, but with an unexplained, much hotter temperature than expected from such a body. 

After re-analysing the \Kepler\ data, van Kerkwijk \etal\ (2010) show that the ellipsoidal variations were themselves asymmetrically modulated, an effect they interpreted as the signature of `Doppler boosting' (also known as the `beaming effect'). Doppler boosting is a relativistic effect of focusing and defocusing the beam of light emitted by a moving source, following its radial velocity variations. A first observation of this effect is reported by Maxted, Marsh \& North (2000). It was recently identified in the \Kepler\ light curve of the binary \object{KPD~1946+4340} (Bloemen \etal\ 2010) and in the \Corot\ light curve of CoRoT-3 (Deleuil \etal\ 2008; Mazeh \& Faigler, submitted). The effect was strong enough in KOI 74 and the light curve precision so good that van~Kerkwijk \etal\ (2010) were able to reconstruct the first radial velocity variations purely from a photometric curve. These authors then derive a mass of $(0.22\pm0.03)$~\Msun\ and argue that the mass and temperature of KOI 74b are compatible with those of a white dwarf. 

However, van Kerkwijk \etal\ (2010) and Rowe \etal\ (2010) had to assume a tidal equilibrium to model the ellipsoidal variations. They used the `equilibrium tide' approximation. According to Pfahl, Arras \& Paxton (2008), this framework works well for stars with masses $<1.4$~\Msun\ and a convective zone. These conditions are clearly not fullfilled by a massive and mostly radiative A1V star like KOI 74. Van Kerkwijk \etal\ (2010) additionally underline that \emph{`[they] cannot solve for the masses using inferences from the lightcurve only'}. The mass derived for the white dwarf from the \Kepler\ photometry could be tested with spectroscopic measurements of the host star's radial-velocity variations.

Velocimetric surveys of early-type stars were started with the HARPS spectrograph at La Silla (Lagrange \etal\ 2009) and SOPHIE spectrograph at the Observatoire de Haute-Provence. So far, these surveys led to the detections of a 25-\Mjup\ brown dwarf around a pulsating 1.7-\Msun\ A9V star (Galland \etal\ 2006), a 9-\Mjup\ planet around a 1.25-\Msun\ F6V star (Galland \etal\ 2005b), a system with two long-period Jupiter-mass planets around a 1.44-\Msun\ F6IV-V star (Desort \etal\ 2008), and a puzzling system around the 1.38-\Msun\ F4V star $\theta$~Cyg (Desort \etal\ 2009). Although KOI 74 is earlier (A1V) and fainter ($V=10.7$~mag) than the bright stars quoted above, the expected amplitude of the radial velocity signal is $\ga 10$~\kms\ (van Kerkwijk \etal\ 2010), well within the capabilities of the SOPHIE spectrograph.

Here, we present measurements of the radial velocity variations of KOI 74, recorded with the SOPHIE spectrograph at the 1.93-m telescope of the Observatoire de Haute-Provence (OHP, France). Because the primary star is a hot and massive early-type star, measuring radial velocities is challenging. The velocimetric follow-up was set by the SOPHIE exoplanet consortium (Bouchy \etal\ 2009) in the frame of the SOPHIE exoplanet sub-programme dedicated to the search for exoplanets around early-type A- and F-type stars. These stars notoriously lack spectral lines, and the few ones available are broadened by the usually high rotation speed. Because of this, a specific pipeline was developed to compute radial velocities for these early-type stars with an accuracy allowing the detection of hot Jupiters (Galland \etal\ 2005a). 

The observations are presented in Sect.~\ref{sec:obs}. We derive and discuss the characteristics of the star KOI 74 and its puzzling companion in Sects.~3 and~4, respectively, and further discuss the white-dwarf nature of KOI~74b in Sect.~5.

\section{Observations and data reduction}
\label{sec:obs}
We observed KOI 74 (\object{KIC 6889235}, \object{2MASS J19531781+4223185}) with the SOPHIE spectrograph at OHP 1.93-m telescope. SOPHIE is a cross-dispersed, environmentally stabilized echelle spectrograph dedicated to high-precision radial velocity measurements from 3\,872 to 6\,943~\AA\ (Perruchot \etal\ 2008; Bouchy \etal\ 2009). Data were recorded using the high-efficiency mode, allowing us to reach a resolving power of $\Delta\lambda/\lambda \approx 40\,000$ while collecting 2.5 times more light than in the high-resolution mode. 

The observations were acquired from March 2010 to early July 2010. Depending on the variable atmospheric conditions, the exposure times ranged between 800 and 1\,800~s, and we obtained signal-to-noise ratios ($S/N$s) per 0.002-nm pixel at 550~nm between 24 and 55. 

Pulsations of the stellar envelope occurring for early-type stars usually result in an excess of scatter on the radial velocity measurements, generally referred to as stellar `jitter'. For an A1V star, the jitter caused by stellar pulsations is expected to dominate other jitter sources such as stellar spots. The jitter can lead to large variations in the stellar radial velocity on typical time scales of a few hours. The technique used to constrain the jitter of early-type stars is to obtain several exposures per night. Each exposure was thus duplicated whenever it was possible (see Sect.~\ref{sec:star}). The total data set includes 23 spectra, and the total exposure time is about 6~h. 

This A1V-star spectrum has few and very broad absorption lines. The SOPHIE pipeline is actually not designed to retrieve a cross-correlation function (CCF) from such a featureless spectrum. We thus derived radial velocities from the files produced by the SOPHIE pipeline with the dedicated software \texttt{SAFIR} (Galland \etal\ 2005a). These images contain the order-by-order spectra resulting from bias subtraction, optimal extraction, cosmic-ray rejection, and division by the spectrum of the flat-field. 

This method consists in correlating, in the Fourier space, each spectrum and a reference spectrum built by summing-up all the available stellar spectra (Chelli 2000). Galland \etal\ (2005a) show that this method can allow planetary to brown-dwarf companions to be found around stars as early as B8, depending on the star properties. 

\begin{figure*}
\resizebox{\textwidth}{!}{\includegraphics{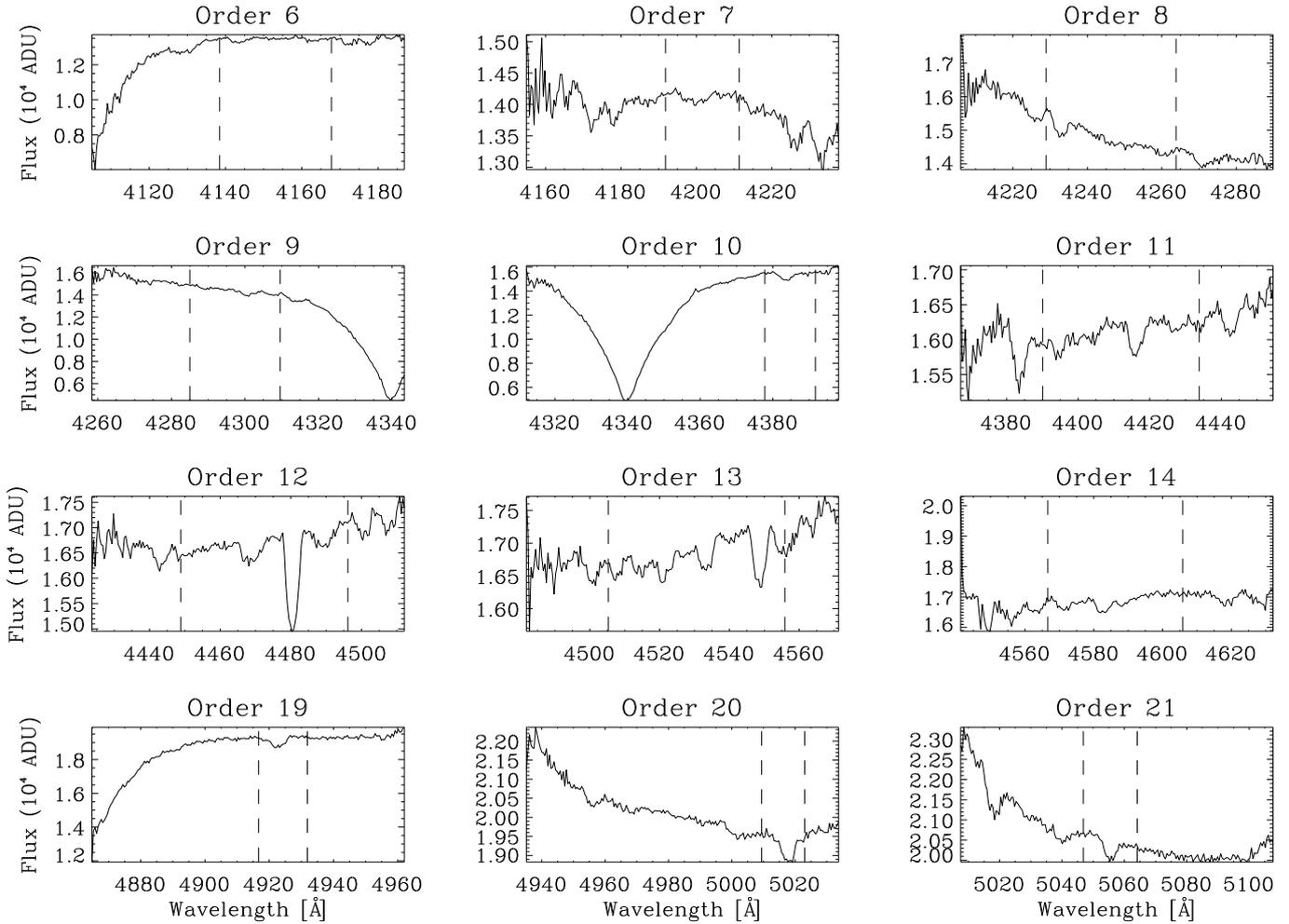}}\caption{\label{fig:spectrum} Mean reference spectrum of KOI 74. Only those spectral orders containing useful information are shown. For each spectral order, the wavelength band used to retrieve the radial velocity of the star is delimited by vertical dashed lines. The spectra have been rebinned by 20.}
\end{figure*}

The first task was to create a mean reference spectrum spectral order by spectral order. The reference spectrum is built in a 2-step process. A first reference spectrum is constructed from the average of all available spectra. The radial velocity of each individual spectrum with respect to this reference is calculated. Each spectrum is then corrected for its velocity shift, and the reference spectrum is finally created from the average of all corrected spectra.

Because the $S/N$ of our observations is $\lesssim 50$ and because the stellar lines are sparse and shallow, most of the orders do not contain useful information. To avoid additional noise caused by intrinsic stellar variations, we excluded the ionized calcium (\ion{Ca}{ii}) H and K lines and the Balmer lines H$\alpha$, H$\beta$, and H$\epsilon$ line (blended with \ion{Ca}{ii} K) from the selection (see Sect.~\ref{sec:star}). We also avoided the spectral regions contaminated by telluric lines. We therefore selected spectral regions in 12 orders (out of 39) with enough stellar lines to recenter each spectrum with respect to the mean reference spectrum. The wavelength bands used to calculate the radial velocities are shown in Fig.~\ref{fig:spectrum}.

The radial velocities of each spectrum were obtained with respect to the reference spectrum. They are reported with the associated uncertainties calculated by \texttt{SAFIR} (see Galland \etal\ 2005a) in Table~\ref{tab:bjd}. The radial velocities are represented as a function of time in Fig.~\ref{fig:rv}. We rejected one measurement from the following analysis because it lies more than 15$\sigma$ away from 2 other points obtained a few minutes apart. This point also has the lowest $S/N$ (24) in the whole series, so we considered it as an outlier. In the following, we deal with 22 measurements.

We caution that in contrast to the `traditional' radial velocities calculated with correlation masks, the radial velocities calculated with the above-described method are \emph{relative to the reference spectrum}. Consequently, they could be different with a different reference, although this would not affect the signal period or amplitude (within the error bars).

\begin{table}
\begin{center}
\caption{\label{tab:bjd} Radial velocities of KOI 74 measured with SOPHIE.}
\begin{tabular}{cccc}
\hline\hline
$\rm BJD-2\,455\,000$ & $S/N$    & $v_r^*$  & $\sigma_{v_r}^{**}$ \\
                      & (550~nm) & (\kms)   & (\kms)         \\
\hline 
268.6844 & 39 & 9.69   & 2.61 \\
268.6940 & 41 & 17.36  & 2.61 \\
271.6725 & 51 & -13.45 & 2.20 \\
271.6813 & 51 & -19.84 & 2.31 \\
273.6664 & 55 & 14.90  & 2.13 \\
273.6739 & 55 & 14.96  & 2.03 \\
291.6380 & 35 & -12.59 & 2.92 \\
291.6488 & 34 & -6.33  & 2.66 \\
297.5922 & 35 & -21.86 & 2.43 \\
327.6063 & 34 & -20.01 & 2.31 \\
336.6112 & 43 & 9.27   & 2.30 \\
367.5910 & 36 & 26.33  & 2.44 \\
367.6006 & 33 & 24.60  & 2.81 \\
368.5914 & 46 & -3.82  & 2.41 \\
368.6022 & 48 & 1.291  & 2.08 \\
379.4241 & 42 & -0.96  & 2.72 \\
379.4337 & 37 & -16.86 & 3.03 \\
379.5866 & 39 & -11.69 & 2.76 \\
382.3863 & 42 & 4.39   & 2.41 \\
382.3937 & 41 & 20.31  & 2.42 \\
383.4697 & 35 & 8.07   & 2.65 \\
383.4771 & 34 & 16.42  & 2.56 \\
\hline
\multicolumn{4}{l}{\parbox{7cm}{
\textbf{Notes.}
$^*$Relative to the mean reference spectrum. 
$^{**}$ The uncertainties given in the last column are calculated with the \texttt{SAFIR} software.
}}
\end{tabular}
\end{center}
\end{table}

\section{Stellar properties of KOI 74}
\label{sec:star}

According to the SIMBAD data base, the star KOI 74 is an early-type A1V star with $V=10.7$~mag. The transit light curve obtained by Rowe \etal\ (2010) allows direct estimation of the star density $\rho_\star$ (Seager \& Mallen-Ornelas 2003), under the assumption that $m_2/m_1 << 1$. The method described by Borucki \etal\ (2010b) for deriving the stellar parameters involves searching a precomputed grid of stellar atmospheres for a best-fit model in the $\rm \{mass,~age,~metallicity\}$ parameter space. This search is constrained by the photometric and spectroscopic measurements of $\rho_\star$ and $T_\mathrm{eff}$, respectively. For KOI 74, Rowe \etal\ (2010) obtained an effective temperature of $9\,400\pm150$~K. Following this `$\rho_\star$ method', they derived a mass, radius, and luminosity of 2.22~\Msun, 1.90~\Rsun, and 25.6~\Lsun, respectively. We adopt the stellar properties derived by Rowe \etal\ (2010), which are listed in Table~\ref{tab:star}.

The distance $d$ of the star can be estimated from the stellar luminosity given above, a bolometric correction $BC=-0.16$ for A1V stars (Strai\v{z}ys \& Kuriliene 1981), and the adapted solar bolometric magnitude ($M_{\mathrm{bol},\sun} = 4.74$) recommended by Torres (2010). We find $\log d = {0.2(V-M_{\mathrm{bol},\sun}+BC)+1}$, hence $d=738$~pc. With the proper motions $\rm \delta_{RA}=3.7$~mas~yr$^{-1}$ and $\rm \delta_{DEC}=-1.0$~mas~yr$^{-1}$ referenced in the CDS, it is possible to retrieve the velocity components of the star on the sky. The barycentric radial velocity of the star is roughly estimated from a Gaussian fit to a not-too-broadened spectral feature, such as the unresolved \ion{Mg}{ii} doublet at 4\,481.06~\AA. (It can be seen in the `order 12' panel of Fig.~\ref{fig:spectrum}.) The spectral shift measured on this line corresponds to a barycentric radial velocity of about $-21$~\kms. The derived Galactic velocity components are then $(U,V,W) \approx (8,-18,-16)$~\kms. According to Leggett (1992), this places KOI~74 in the kinematic young disc of the Galaxy, with a typical age of $\sim600$~Myr.

We derive a stellar projected rotational velocity $v_\mathrm{rot} \sin i_\star \approx 142\pm5$~\kms. This value is obtained through an autocorrelation of the reference spectrum, and is consistent with the value of $150$~\kms\ measured by van Kerkwijk \etal\ (2010) on a Keck~I/HIRES spectrum. Given the stellar radius of 1.90~\Rsun, this would yield a projected rotation period $P/\sin i_\star \approx 0.7$~day. This value is in good agreement with the period inferred by van Kerkwijk \etal\ (2010) from a weak modulation period in the \Kepler\ light curve. Since the rotation period derived by photometry is compatible with the spectroscopic projected rotation period, we should have $\sin i_\star \sim 1$. Therefore, it seems reasonable to assume that the stellar and orbital spins are aligned and inclined by $i_\star \approx i=88.8\degr$. Furthermore, a rotation period of around 0.7~day is also suggested by the periodogram of the radial velocity measurements (see Sect.~\ref{sec:wd}).

The value derived for $v_\mathrm{rot} \sin i_\star$ is lower than the typical value ($\sim 200$~\kms) measured for other A stars followed by the radial velocity survey described in Lagrange \etal\ (2009). We have argued above that $\sin i_\star \sim 1$, so the low value of $v_\mathrm{rot}$ cannot be a projection effect. We surmise that the stellar rotation could have slowed down as a result of gravitational interactions with the massive companion (e.g., tidal effects subsequent to mass transfer).

Some spectral indicators of stellar activity are presented in Fig.~\ref{fig:CaHK}. No emission is seen in the cores of the broad \ion{Ca}{ii} H and K lines (Fig.~3 top panel), whereas a thin interstellar absorption is observed in the \ion{Ca}{ii} K line. (The \ion{Ca}{ii} H line is blended with the Balmer H$\varepsilon$ line). This suggests a lack of chromospheric activity consistent with the stellar spectral type. 
Intriguing emissions are seen in the H$\alpha$ (Fig.~\ref{fig:CaHK} middle panel) and H$\beta$ lines of KOI 74. As can be seen in Fig.~\ref{fig:CaHK} (bottom panel), these emissions are also detected in the spectrum recorded by SOPHIE's second fiber  located on the sky, 2\arcmin -away from the star. This shows that the Balmer emissions seen in KOI~74 spectra come mainly from a sky contamination. The shift between the Balmer emissions and the broad stellar Balmer absorptions varies because of the reflex motion of KOI~74 with respect to the sky.

\begin{figure}
\resizebox{\columnwidth}{!}{\includegraphics{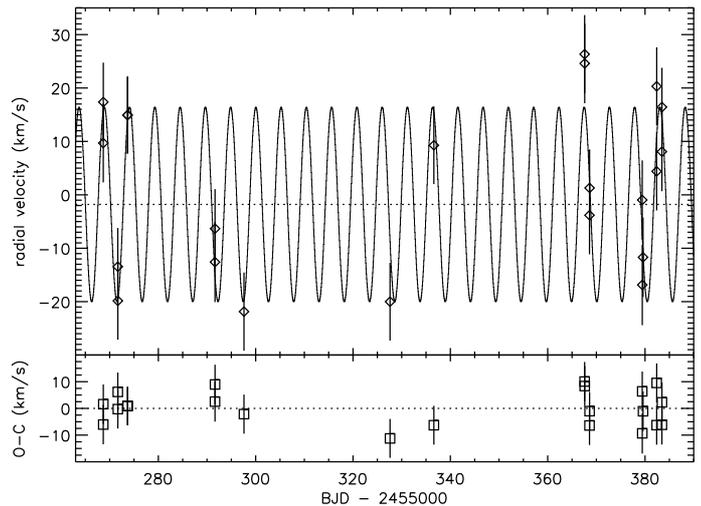}}\caption{\label{fig:rv}\emph{Top:} Radial velocity measurements (diamonds) of KOI 74 as a function of time. The error bars include the jitter dispersion. The black line is the best Keplerian fit to the measurements. \emph{Bottom:} Observed minus calculated residuals (squares).}
\end{figure}

\begin{figure}
\resizebox{\columnwidth}{!}{\includegraphics{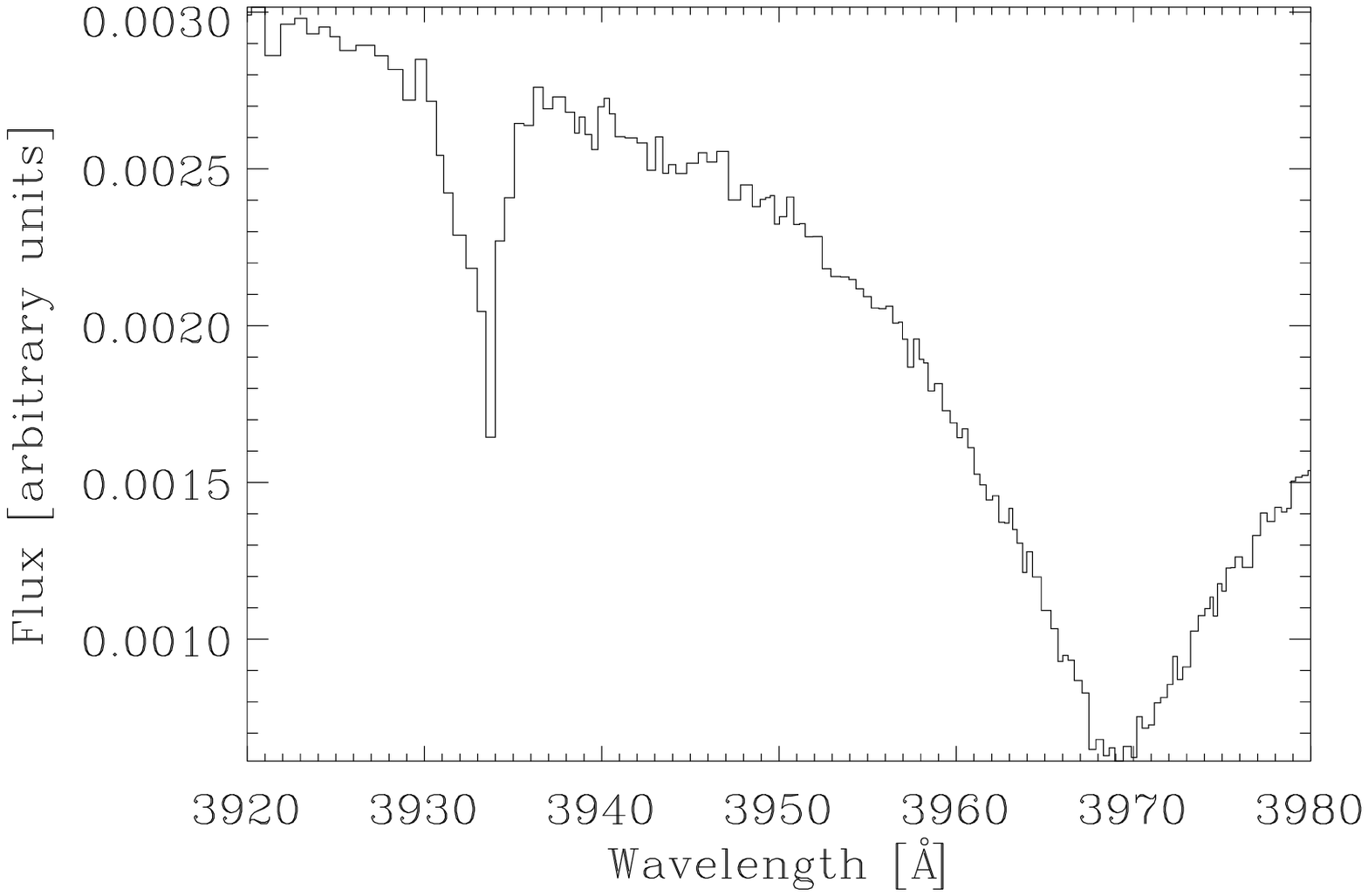}}
\resizebox{\columnwidth}{!}{\includegraphics{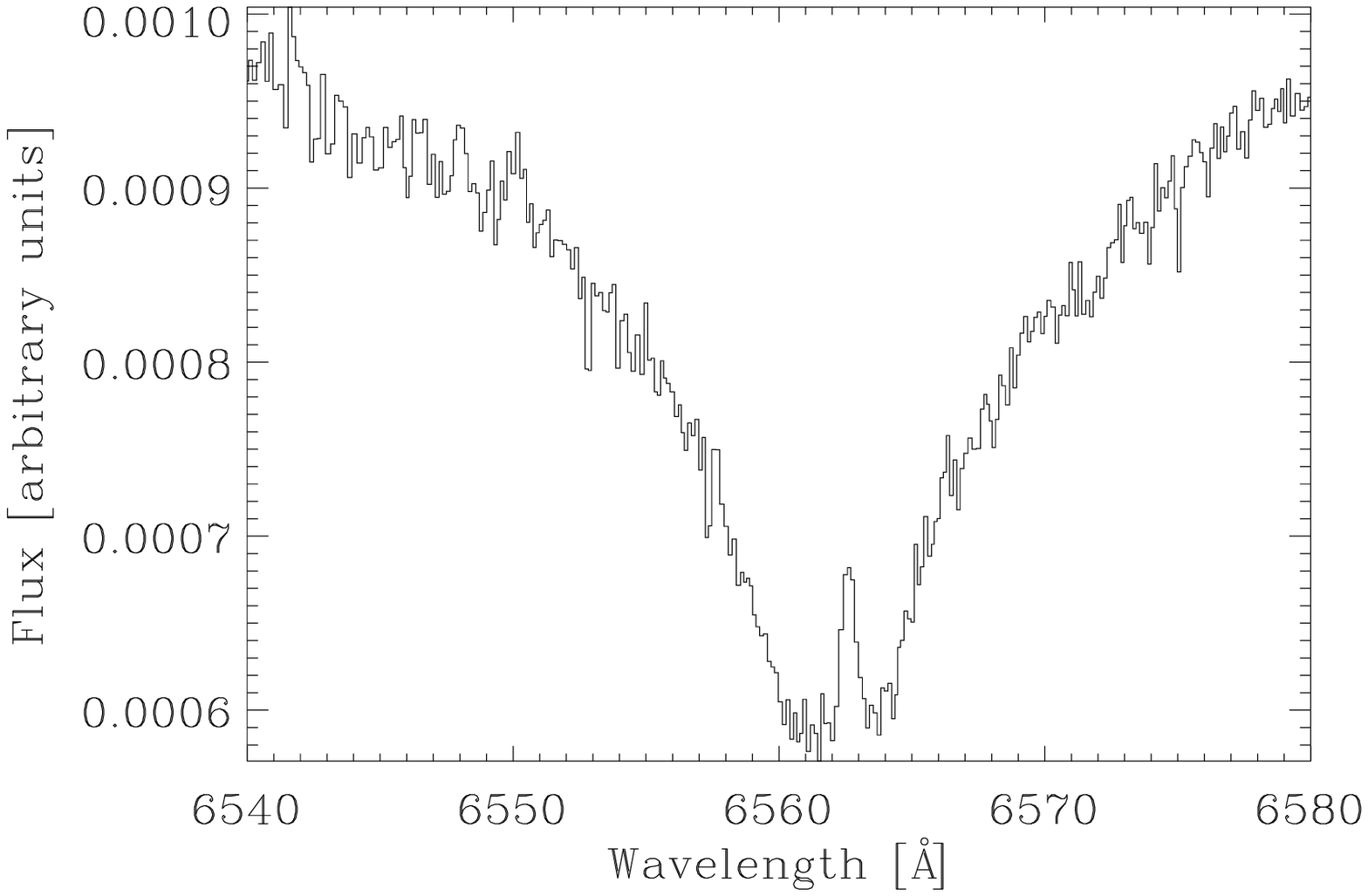}}
\resizebox{\columnwidth}{!}{\includegraphics{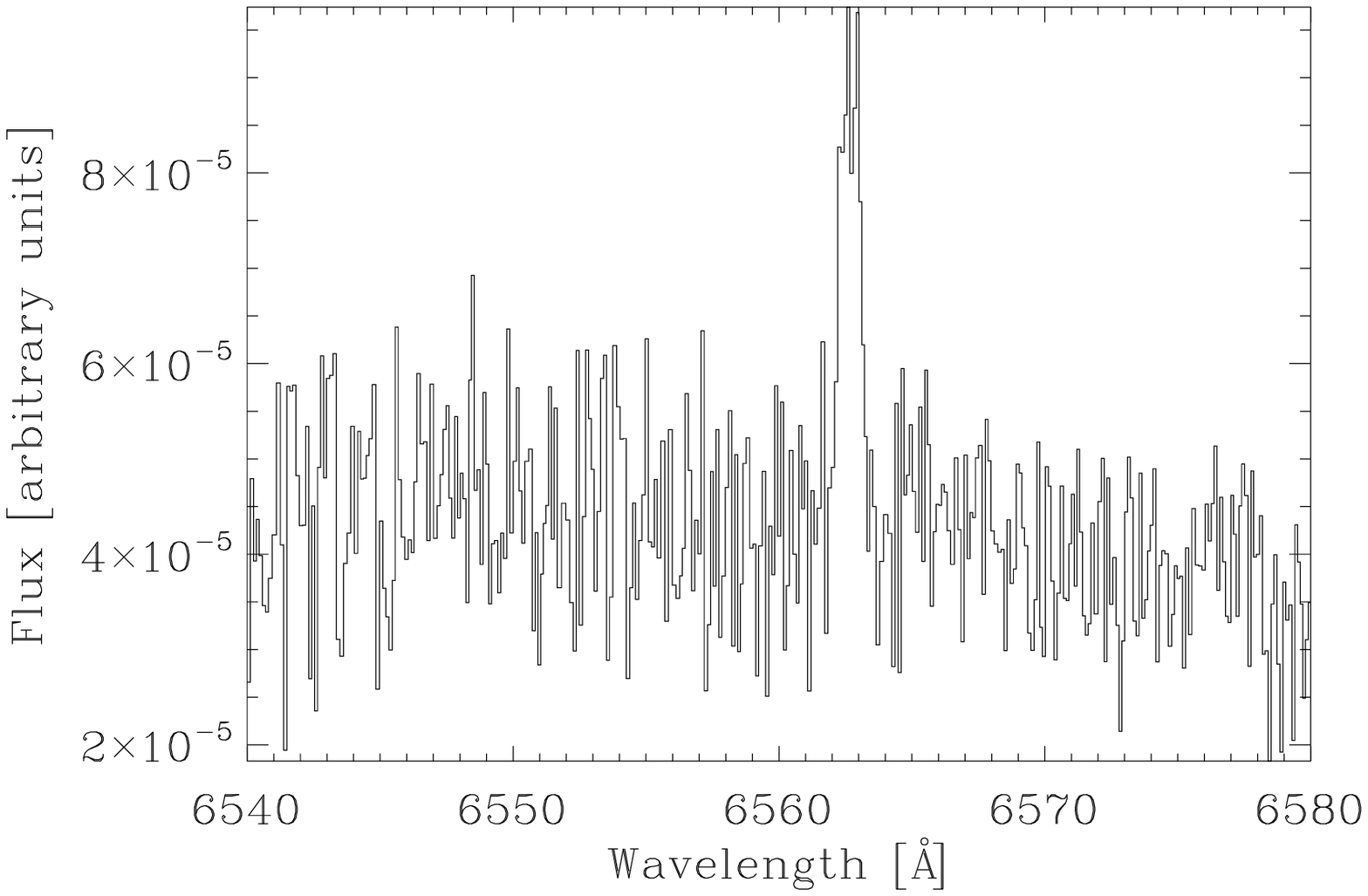}}
\caption{\label{fig:CaHK} \emph{Top:} The \ion{Ca}{II} H line at 3\,968~\AA\ (blended with H$\varepsilon$) and \ion{Ca}{II} K line at 3\,933~\AA\ of KOI 74, which features a thin interstellar absorption. \emph{Middle:} The H$\alpha$ line at 6\,563~\AA. \emph{Bottom:} Sky spectrum recorded with SOPHIE's second fiber (2\arcmin\ from KOI 74) in the region of the H$\alpha$ line. The spectra have been rebinned by 10. }
\end{figure}

To constrain the stellar jitter, we acquired -- whenever possible -- two consecutive exposures of KOI 74, separated by a few minutes. In 1~min, the maximum variation in the star's radial velocity due to its companion is close to $4 K_1 / P \sim 10$~\ms, where $K_1$ is the semi-amplitude of the radial velocity variations and $P$ the period (expressed in minutes). Such variations are not detectable given the precision achieved within this data set. Consequently, any detected variations observed between consecutive exposures are due to the stellar jitter. Our measurements show that the star radial velocity can vary by as much as $\sim 15$~\kms\ in less than 20~min. The dispersion among points taken consecutively on one night is 6.9~\kms. We quadratically added this `jitter dispersion' to the errors listed in Table~\ref{tab:bjd}. We obtained the error bars shown for the data points in Fig.~\ref{fig:rv}.

\begin{table}
\begin{center}
\caption{\label{tab:star} Adopted stellar parameters for KOI 74.}
\begin{tabular}{lcc}
\hline\hline
Parameter                            & Value                     & Ref.\\
\hline
$V$ (mag)                            & 10.7                      & 1 \\
Spectral type                        & A1V                       & 1 \\
$B-V$ (mag)                          & 0.08                      & 1 \\
$v_\mathrm{rot} \sin i_\star$ (\kms) & $142\pm5$                 & 2 \\
$T_\mathrm{eff}$ (K)                 & $9\,400 \pm 150$          & 3 \\
Mass (\Msun)                         & $2.22^{+0.10}_{-0.14}$    & 3 \\
Radius (\Rsun)                       & $1.899^{+0.043}_{-0.051}$ & 3 \\
Luminosity (\Lsun)                   & $25.6\pm2.4$              & 3 \\     
\hline
\multicolumn{3}{l}{\parbox{6cm}{
\textbf{References.} (1) CDS; (2) this work; (3) Rowe \etal\ 2010.
}}
\end{tabular}
\end{center}
\end{table}

\begin{figure}
\resizebox{\columnwidth}{!}{\includegraphics{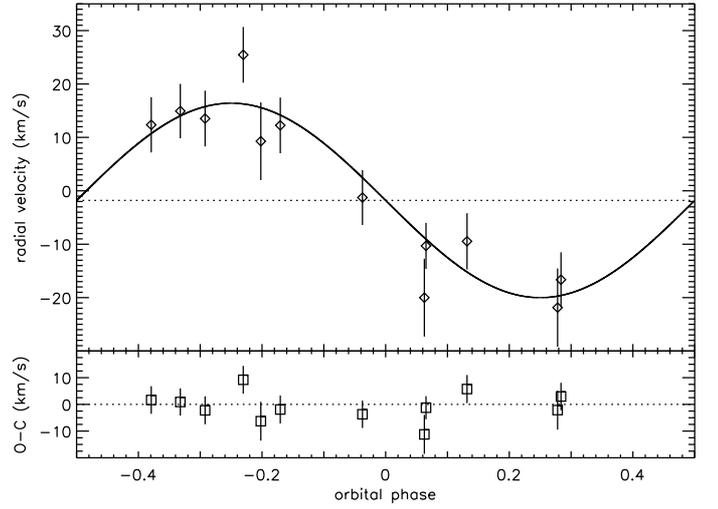}}
\caption{\label{fig:phase}\emph{Top:} SOPHIE phase-folded radial velocities (diamonds). Each point and its error bar (including the jitter dispersion) result from the average of all points obtained on the same night. The black line is the best Keplerian fit to the data. \emph{Bottom:} Observed minus calculated residuals (squares).}
\end{figure}

\begin{figure}
\resizebox{\columnwidth}{!}{\includegraphics{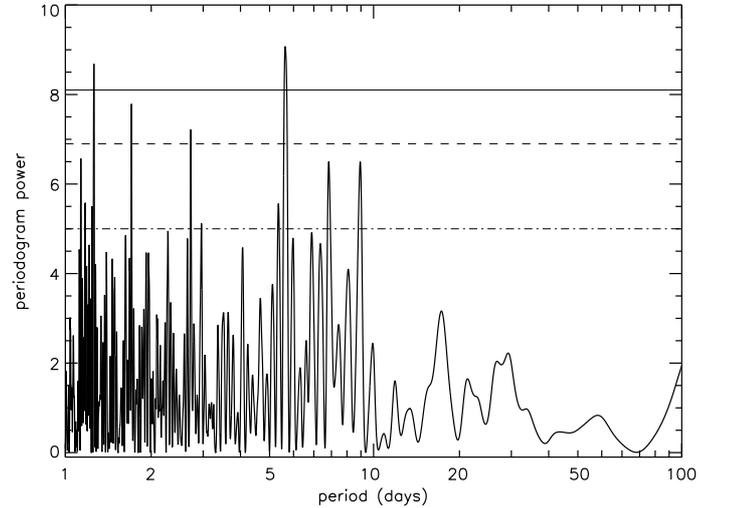}}
\caption{\label{fig:period}Periodogram of the velocimetric measurements presented in Table~\ref{tab:bjd}. The false-alarm probability (FAP) levels are reported as horizontal lines, indicating the significance of the power peaks: $1\sigma$ (dash-dotted line), $2\sigma$ (dashed line), and $3\sigma$ (plain line).}
\end{figure}

\section{\textbf{Velocimetric} detection of a massive companion}
\label{sec:wd}

The SOPHIE radial velocities of KOI 74 are plotted in Fig.~\ref{fig:rv} as a function of time. They are phase-folded on the \Kepler\ period derived by Rowe \etal\ (2010) in Fig.~\ref{fig:phase}. They cover a period of about 4~months and show peak-to-peak variations as large as 40~\kms, with a standard deviation of 15.4~\kms, well above the dispersion due to the jitter (6.9~\kms). The typical uncertainty on an individual measurement is 2.5~\kms, better than the stellar jitter. 

Figure~\ref{fig:period} shows the periodogram of the radial velocities. There is a clear peak at about $5.2$~days, matching with the \Kepler\ period of 5.19~days. To estimate the false-alarm probability (FAP) for that peak, we used a
bootstrap resampling of the true measurements, and examined the peak power in the periodogram of each of these virtual data sets. Among 10\,000 trials, only three are found with a peak higher than the peak detected for the true $\sim5.2$-day signal, suggesting an FAP $\lesssim 3\times10^{-4}$. We computed the levels for $\rm FAP=99.7\%$ (3$\sigma$), 95.4\% (2$\sigma$), and 68.3\% (1$\sigma$) and reported them in Fig.~\ref{fig:period}. There is another peak standing above the 99.7\% level at a period of 1.24 days. It is the alias of the 5.2-day period with $\sim1$-day sampling. Two other peaks at 1.64 and 2.57~days are detected above the 2-$\sigma$ confidence level: these frequencies are the aliases of a 0.72-day period for a 1-day sampling. This 0.72-day period is compatible with the rotation period of the star according to van Kerkwijk \etal\ (2010) and our own measurement of $v_\mathrm{rot} \sin i_\star$.

The power peak at the \Kepler\ period suggests that the companion signal has been detected. Indeed, the subtraction of a Keplerian model to the data using the \Kepler\ ephemeris significantly improves the dispersion of the $\rm observed - calculated$ ($\rm O-C$) residuals. We fixed the period $P$ and time of mid-transit ($T_0$) to the \Kepler\ values calculated by Rowe \etal\ (2010). These authors detected 8 transits and 9 occultations of the companion. Since the occultations always happen $\sim P/2$ after the transits, the orbit is probably circular, unless the longitude of the periastron with respect to the line of sight is 0\degr\ or 180\degr. Besides, a massive body on a 5-day period around a massive star is expected to be circularized. Van Kerkwijk \etal\ (2010) set $e < 0.03$ as an upper limit for the eccentricity. In the following, we consider that $e=0$.

The search for a Keplerian fit with 2 free parameters (the semi-amplitude $K_1$ and the systemic velocity $\gamma$) to 22 measurements yields a best-fit solution with $K_1 = 18.2\pm1.7$~\kms\ and $\gamma = -1.8\pm1.4$~\kms. The systemic velocity is expected to be $\sim0$~\kms\ since \texttt{SAFIR} radial velocities are calculated with respect to the mean of all spectra. The dispersion of the residuals ($\rm O-C$) is 6.1~\kms, close to the dispersion due to the stellar jitter (6.9~\kms) that dominates the photon noise (2.5~\kms). The agreement between the $\rm O-C$ residual and jitter dispersions ensures that our Keplerian model accurately reproduces the radial velocity variations.

Finally, we averaged all points obtained during one night to level the short-term variability. Twelve points remain, phase-folded in Fig.~\ref{fig:phase}. A Keplerian fit gives similar results to those quoted above. The dispersion of the residuals decreases to 4.7~\kms.

In Table~\ref{tab:wd}, we report the orbital and physical properties of the companion. Assuming a stellar mass $m_1 = 2.22\pm0.14$~\Msun\  and $\sin i \approx 1$ (Rowe \etal\ 2010), we found a mass for the companion of $m_2 = 264\pm26$~\Mjup\ or $0.252\pm0.025$~\Msun. 

\section{The low-mass transiting white dwarf KOI 74b}

The transits and occultations observed with \Kepler\ allowed Rowe \etal\ (2010) and van Kerkwijk \etal\ (2010) to estimate the companion radius $r_2 = 0.04$~\Rsun\ and brightness temperature $T_2 \approx 13\,000\pm1\,300$~K, respectively (assuming a 10\% uncertainty for the temperature; van Kerkwijk \etal\ 2010). We can also derive a mean density $\rho = 5.86$~kg~cm$^{-3}$ and $\log g=6.65$. These values are in good agreement with what is known for low-mass white dwarfs in binary systems. 

In fact, while the majority of white dwarfs, with an average mass around 0.6~\Msun, evolve from main-sequence stars following normal evolutionary processes, extremely low-mass white dwarfs ($\la 0.2$~\Msun) are only observed in binary systems and no single-star evolution path is known that could form these objects (Kawka \& Vennes 2009).

\begin{figure*}
\resizebox{\textwidth}{!}{\includegraphics{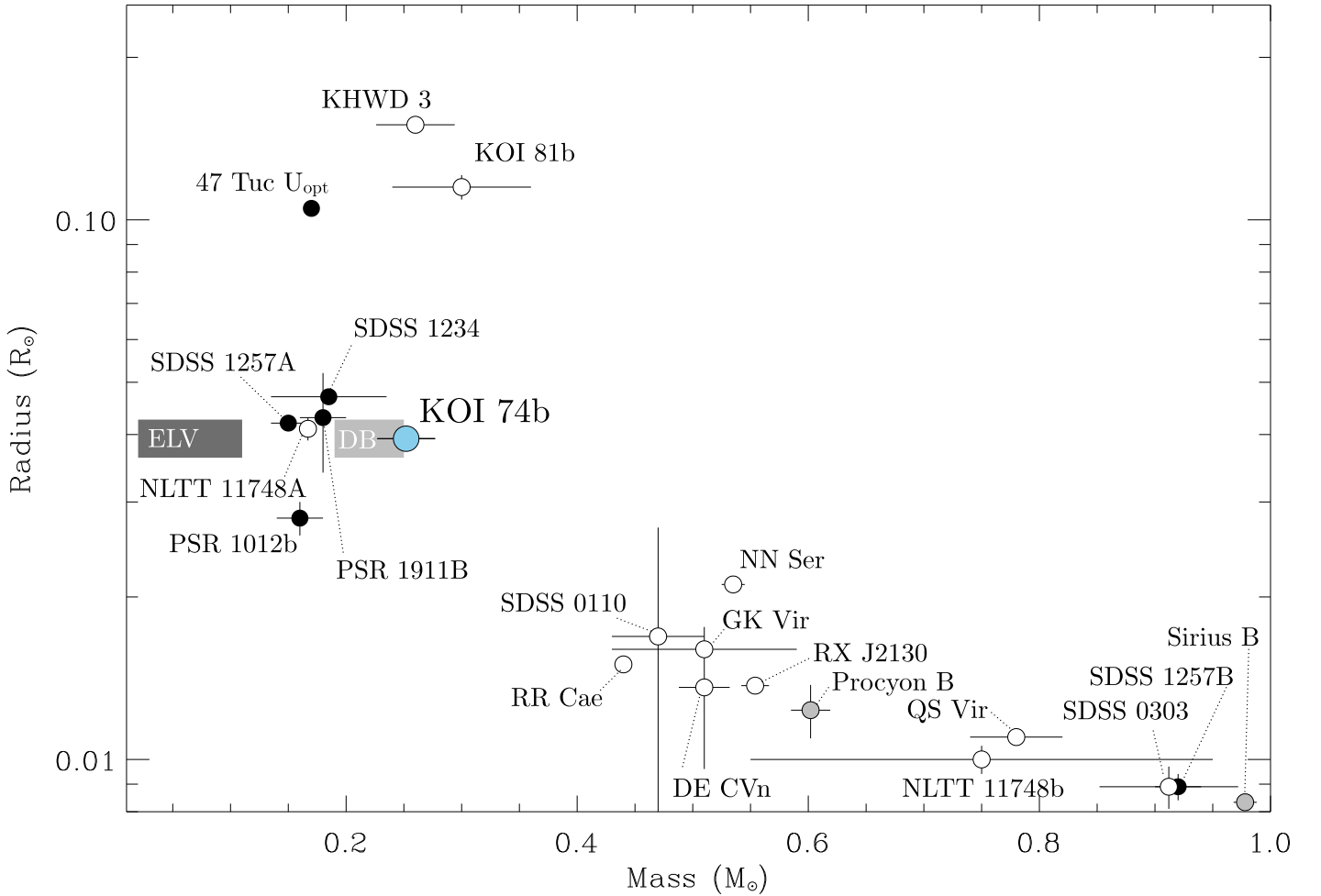}}\caption{\label{fig:wd} Mass-radius diagram in which KOI 74b (sky blue symbol) is represented with other white dwarfs in detached or post-common-envelope eclipsing binaries (open symbols) and noneclipsing low-mass white dwarfs (black-filled symbols). The mass ranges for KOI 74b previously calculated with ellipsoidal light variations (ELV; Rowe \etal\ 2010) and Doppler boosting (DB; van Kerkwijk \etal\ 2010) are indicated by thick dark and light gray error bars. The mass of KOI~81b is taken from van Kerkwijk \etal\ (2010) with an uncertainty of 20\%. The mass indicated for KHWD~3 is the value obtained by Carter, Rappaport \& Fabrycky (submitted) with Doppler boosting. Procyon B (Provencal \etal\ 2002) and Sirius B (Barstow \etal\ 2005) are shown for comparison (gray-filled symbols). References for the other objects are given in the text.}
\end{figure*}

In Fig.~\ref{fig:wd}, we reported the masses and radii of several white dwarfs from the literature for comparison purposes with KOI 74b. We included a sample of white dwarfs in post-common-envelope eclipsing binaries (PCEEB) (Parsons \etal\ 2010, and references therein). The masses and radii of these degenerate stars can be directly derived with velocimetry and photometry, similarly to KOI 74b. Few of these eclipsing `detached'\footnote{Where mass transfer can occur but is dynamically unstable, contrary to cataclysmic variables with stable mass transfer between both components.} systems are known. Most of them are paired with an M dwarf that is less massive than the white dwarf itself. As can be seen in Fig.~\ref{fig:wd}, KOI 74b is both larger and less massive, i.e., less dense, than any other PCEEB white dwarf known so far. This might point to a difference in internal composition or to a different evolutionary path since the primary star KOI~74 is much more massive than the white dwarf companion. This agrees with the discussion of van Kerkwijk \etal\ (2010) about the unlikely common-envelope evolution of the KOI 74 system.

In fact, KOI 74b seems more similar to extremely low-mass (ELM) white dwarfs. These objects were initially detected as companions to pulsars, such as \object{PSR~J1012+5307b} (van Kerkwijk, Bergeron \& Kulkarni 1996), \object{PSR~J1911-5958b} (Bassa \etal\ 2006), \object{47~Tuc~U} (Edmonds \etal\ 2001), or \object{SDSS~1257+5428} (Kulkarni \& van Kerkwijk 2010). Some ELM white dwarfs identified in deep surveys are known as binaries solely on the basis of radial velocity variations, since their companions remain unseen (see for instance the case of \object{SDSS~J123410.37-022802.9}; Liebert \etal\ 2004). (The radii of these objects are usually determined from mass-radius relations.) Recently, the \object{NLTT~11748} system (Kawka \& Vennes 2009; Kawka, Vennes \& Vaccaro 2010) has been identified as the first ELM-white dwarf eclipsing binary (Steinfadt \etal\ 2010), allowing a direct estimation of an ELM radius. On the other hand, the KOI 74 system again distinguishes itself from these systems, owing to the mass of the primary A1 star.

Van Kerkwijk \etal\ (2010) note that the KOI 74 system, and maybe KOI 81, could be downscale versions of the Regulus binary system, consisting in a 3.4-\Msun\ B7V star ($\alpha$~Leo~A) and a recently identified $\sim0.3$~\Msun\ white dwarf ($\alpha$~Leo~Ab) on a 40-day orbit (Gies \etal\ 2008). Besides, Di Stefano (submitted) proposes that the primary stars KOI 74 and KOI 81 could be classified as blue stragglers: \emph{`A viable hypothesis is that [the white dwarfs] are the cores of stars that have each been eroded or disrupted by a companion star. The companion, which is the star monitored today, is likely to have gained mass from its now-defunct partner, and can be considered to be a blue straggler.'} In fact, we found that the blue straggler system F190 in the M67 open cluster (Milone 1991; Milone \& Latham 1992) is a close analog to the KOI 74 system, with component masses of 0.2 and 2.1~\Msun\ and with a period of 4.18~days. (This system is, however, eccentric.\footnote{This puzzling eccentricity could result from a close encounter with a third body as the system is  part of an open cluster (D.~Latham, private communication); other possible reasons have been discussed by Milone \& Latham (1992).}) 

To our knowledge, systems composed of an extremely low-mass white dwarf or a low-mass white dwarf and an early-type main-sequence star are excessively rare. Meanwhile, the rarity of these systems may not be intrinsic but could result from the following bias: in imaging surveys such as the Sloan Digital Sky Survey, the detected white dwarf-main sequence star binaries usually include a late type (M or K) dwarf, because such a cool star does not outshine the white dwarf. Nevertheless, the white dwarf mass distribution inferred for these binary systems peaks at 0.5~\Msun, close to the average mass of single white dwarfs (Rebassa-Mansergas \etal\ 2010). 

Owing to its rarity, a system like KOI 74 will be an invaluable target for testing atmospheric models commonly used to derive radii and masses for white dwarfs (e.g., Serenelli \etal\ 2001, 2002) and binary evolution scenarios. 

\begin{table}
\begin{center}
\caption{\label{tab:wd} Parameters for KOI 74b and its orbit.}
\begin{tabular}{lr}
\hline\hline
Parameter                           & Value                    \\
\hline
$P$ (days)                          & $5.188754^*$ (fixed)         \\
$T_0$ (BJD)                         & $2\,454\,958.8799^*$ (fixed) \\
$e$                                 & 0 (fixed)                \\
$K_1$ (\kms)                        & $18.2\pm1.7$             \\
$\gamma$ (\kms)                     & $-1.8\pm1.4$             \\
$\rm \sigma(O-C)$ (\kms)            & $4.7$                     \\
$i$ (\degr)                         & $88.8^*$                   \\
$m_2$ (\Mjup)                       & $264\pm26$               \\
$m_2$ (\Msun)                       & $0.252\pm0.025$          \\
$r_2$ (\Rsun)                       & $0.0393\pm0.0013^*$        \\
$\rho_2$ (kg~cm$^{-3}$)             & 5.86 \\
$\log g$ (cgs)                      & 6.65 \\
\hline   
\multicolumn{2}{l}{\parbox{5cm}{\textbf{Notes.} $^*$Value from Rowe \etal\ 2010}}
\end{tabular}
\end{center}
\end{table}

\section{Conclusion}

We have obtained radial velocity measurements of KOI 74, a massive and early main-sequence star transited by a hot compact object. The measurements indicate that the companion is a 0.25-\Msun\ object, most probably a low-mass white dwarf. We emphasize that this transiting low-mass white dwarf, whose mass, radius, and temperature have been directly measured through radial velocimetry and photometry, will be a precious template object for binary evolution theories.

This work provides an independent validation of the approach of van Kerkwijk \etal\ (2010), who measured a radial velocity curve from a high-precision photometry using the relativistic Doppler boosting effect. Consequently, both the spectroscopic and Doppler boosting methods can be used to estimate the radial velocity amplitude of stars with massive companions (see also Bloemen \etal\ 2010). Carter, Rappaport \& Fabrycky (submitted) recently announced a third transiting white dwarf detected with \Kepler. In this case, the mass ratio between the white dwarf and its A-type companion is found to be inconsistent when calculated with ellipsoidal light variations or Doppler boosting. The present work clearly favours the Doppler boosting estimation.

This technique will be, however, limited to those binaries where the companion is massive enough to trigger large radial velocity variations and a detectable Doppler boosting effect. In the planet-to-brown-dwarf domain, an encouraging detection has recently been performed for the planet CoRoT-3b (22~\Mjup) by Mazeh \& Faigler (submitted). 

Spectroscopic velocimetry will remain a much needed method of establishing the true nature of transit candidates and determining their masses. Using the SOPHIE spectrograph and a 1.93-m telescope, we have been able to measure the radial velocity variations of a rapidly rotating and early-type star of magnitude $V=10.7$. This is a promising debut for the follow-up of \Kepler\ targets.

\begin{acknowledgements}
The authors would like to thank the referee Marten van Kerkwijk for a helpful review. We are also grateful to David Latham, Jason Rowe, Lars Buchhave, Adela Kawka, Avi Shporer, and Josh Carter for stimulating discussions. D.E.\ is supported by the Centre National d'\'Etudes Spatiales (CNES). A.E.\ is supported by a fellowship for advanced researchers from the Swiss National Science Foundation (grant PA00P2\_126150/1). N.C.S.\ acknowledges the support by the European Research Council/European Community under the FP7 through a Starting Grant, as well as in the form of grants reference PTDC/CTE-AST/66643/2006 and PTDC/CTE-AST/098528/2008, funded by Funda\c{c}\~ao para a Ci\^encia e a Tecnologia (FCT), Portugal. N.C.S.\ would further like to thank the FCT for support through a Ci\^encia\,2007 contract funded by FCT/MCTES (Portugal) and POPH/FSE (EC). These results have made use of the SIMBAD and VizieR data bases, operated at the CDS, Strasbourg, France.
\end{acknowledgements}

\end{document}